\documentclass{elsart}

\usepackage{latexsym}
\usepackage[latin1]{inputenc}       
\usepackage[english]{babel}        
\usepackage[dvips]{graphicx}
\usepackage{psfrag}

\usepackage{amsbsy} 
\usepackage{subfigure}

\usepackage{pstricks}

\newcommand{\vect}[1]{\boldsymbol{#1}} 
\newcommand{\de}{\mathrm{d}}        

\newcommand{\px}{P{\textsc{hotonics}}}
\newcommand{\cpp}{\mbox{\textsc{c}}{\tiny\raisebox{.5ex}{++}}}

\newcommand{\iA}{\mathrm{A}}
\newcommand{\iB}{\mathrm{B}}
\newcommand{\iAB}{\mathrm{AB}}

\newcommand{\eff}{\mathrm{e}}
\newcommand{\abb}{\mathrm{a}}
\newcommand{\sct}{\mathrm{s}}

\newcommand{\ngr}{n_\mathrm{\mathrm{g}}}
\newcommand{\nph}{n_\mathrm{\mathrm{p}}}

\newcommand{\src}{\mathrm{s}}
\newcommand{\rec}{\mathrm{rec}}

\newcommand{\ppdf}{\mathrm{pdf}}

\newrgbcolor{rgbred}{1  0.0   0.0}
\newrgbcolor{rgbdred}{0.4  0.0   0.0}
\newrgbcolor{rgbblack}{0.0  0.0   0.0}

\providecommand{\chv}{Cherenkov }
\providecommand{\comment}[1]{~}

\providecommand{\degr}{\mbox{$^\circ$}}

\providecommand{\usp}{\hspace*{0.2em}}
\providecommand{\amanda}{\textsc{Amanda~}}

\usepackage{amssymb}

\begin{document}

%
%
%
%

\begin{frontmatter}

\title{Light tracking through ice and water -- Scattering and
absorption in heterogeneous media with P\normalsize{HOTONICS} }

\author[Uppsala]{J. Lundberg\corauthref{cor}},
 \corauth[cor]{Corresponding author\\
Phone:~+4670-7422777\\
Fax: +4618-4713513 }
\ead{johan.lundberg@tsl.uu.se}\ead[url]{photonics.tsl.uu.se}
\author[Pedloc]{P. Mio\v{c}inovi\'c},
\author[kurtloc]{K. Woschnagg},
\author[Burgloc]{T. Burgess},
\author[adamsloc]{J. Adams},
\author[Burgloc]{S. Hundertmark},
\author[paulloc]{P. Desiati},
\author[nissloc]{P. Niessen}
\address[Uppsala]{\mbox{Division of High Energy Physics, Uppsala University,  Uppsala, SE}}
\address[Pedloc]{\mbox{Department of Physics and Astronomy, University of Hawaii,  Manoa, US}}
\address[kurtloc]{\mbox{Department of Physics, University of California, Berkeley, California, US}}
\address[Burgloc]{\mbox{ Department of Physics, Stockholm University, Stockholm, SE}}
\address[adamsloc]{\mbox{Department of Physics and Astronomy, University of Canterbury, Christchurch, NZ}}
\address[paulloc]{\mbox{Department of Physics, University of Wisconsin, Madison, Wisconsin, US}}
\address[nissloc]{\mbox{Bartol Research Institute, University of Delaware, Newark, Delaware, US}}

\begin{abstract}

In the field of neutrino astronomy, large volumes of optically
transparent matter like glacial ice, lake water, or deep ocean water are
used as detector media. Elementary particle interactions are studied
using in situ detectors recording time distributions and fluxes of the
faint photon fields of \chv radiation generated by ultra-relativistic
charged particles, typically muons or electrons.

The \px{} software package was developed to determine photon flux and
time distributions throughout a volume containing a light source through
Monte Carlo simulation. Photons are propagated and time distributions
are recorded throughout a cellular grid constituting the simulation
volume, and Mie scattering and absorption are realised using wavelength and position
dependent parameterisations. The photon tracking results are stored in
binary tables for transparent access through \mbox{\textsc{ansi-c}} and
\cpp{} interfaces. For higher-level physics applications, like
simulation or reconstruction of particle events, it is then possible to
quickly acquire the light yield and time distributions for a
pre-specified set of light source and detector properties and
geometries without real-time photon propagation.

In this paper the \px{} light propagation routines and methodology are
presented and applied to the IceCube and \textsc{Antares} neutrino
telescopes. The way in which inhomogeneities of the Antarctic glacial
ice distort the signatures of elementary particle interactions, and how
\px{} can be used to account for these effects, is described.

\end{abstract}
\end{frontmatter}


\pagebreak

\section{Introduction}


In optical high energy neutrino astronomy light from particle physics
events is observed using a large number of detectors placed deep in
glacial ice or in ocean or lake water. Successful simulation and
reconstruction of such events relies on accurate knowledge of light
propagation within the detector medium. Light propagating through even
the clearest water or ice is affected by scattering and absorption. For
light sources and receivers separated by distances comparable to the
photon mean free path, scattering effects can neither be analytically
calculated nor ignored. The typical scattering lengths in these
detection media are tens to hundreds of metres. Since this scale is
comparable to the typical detector separation, detailed simulation of
the photon propagation is required to obtain information necessary for
event simulation and reconstruction. The problem is complicated further
by the anisotropy of the light emitted in particle interactions and the
heterogeneity of detector media. \px{} is a freely available software
package~\cite{pxcode} containing routines for detailed photon Monte
Carlo simulations, which take into account such complexities to provide,
in tabulated form, the photon flux distribution throughout a specified medium for an input light
source.

With \px{} the photon flux and time distributions can be tabulated for
an arbitrarily large volume of a propagation medium, for a user defined
range of light source and detector properties. This means that once a
\px{} table set has been generated for a class of light sources and
detectors, it is possible to quickly and transparently acquire the light
yield and time distributions without any need for real-time photon
propagation during, for example, particle physics event simulation or
reconstruction. This is made possible by the \px{} table reader library,
with which a user (program) can dynamically query the pre-calculated
tables by specifying the locations and geometrical relations between
light sources and detectors.
A simulation chain for a complete experimental setup can be achieved by
using these interfaces and applying detector specific details such as
modelling of electronics, data acquisition, and triggers. For event
reconstruction, \px{} provides probability density functions for
arrival times of
independent photons and the expected number of detected photons.

In this paper we first introduce the relevant physics of the photon
propagation (Section~\ref{diftheory}) and the details of the \px{}
implementation (Section~\ref{photomc}). We then compare our results with
observations of calibration light sources in sea water and glacial ice
(Section~\ref{obscomparison}). In Section~\ref{neuastro}, we present
some photon tracking results relevant to the detection of neutrinos with
the IceCube neutrino telescope.

\section{Light propagation in diffuse media}

\label{diftheory}

Our goal was to model the transport of light through glacial ice and
water. Photon propagation depends on the optical properties of the
medium, in particular on the velocity of light and the absorption and
scattering cross sections. Glacial ice is optically inhomogeneous because of
depth dependent variations in temperature, pressure, and concentrations
of air bubbles and insoluble dust. Since the dust deposits track
climatological changes and are therefore assumed to be arranged in
horizontal layers, their effect is parameterised as a vertical variation
of the optical properties. In addition to this spatial variation, the
wavelength dependence of the medium parameters must be taken into
account.
Before describing the implementation to achieve our goal,
we review the optical quantities that must be considered
in the simulation, using notation in which the wavelength dependence is
left implicit.

The time of light travel is determined by the group velocity of light,
which is given by the group refractive index $\ngr$, while
various transmission and scattering coefficients depend on the phase
velocity~\cite{pwveloc} and its index of refraction $\nph$.

Absorption of visible and near UV photons in pure water and ice is due
to electronic and molecular excitation processes and is characterised by
the \emph {absorption length} $\lambda_{\abb}$. Measurements of light
attenuation have been performed at relevant wavelengths in lake water by
the Baikal~\cite{baikal} collaboration, and in sea water by the
\textsc{Dumand}~\cite{dumref}, \textsc{Nestor}~\cite{nestor},
\textsc{Antares}~\cite{antareswtr}, and \textsc{Nemo}~\cite{nemoref}
 collaborations. The
\amanda collaboration has developed an empirical model for optical
absorption in deep glacial ice by combining laboratory and in situ
measurements~\cite{amandaice}.

Photon scattering by scattering centres of general sizes is described by
Mie scattering theory~\cite{mietheory}, which for any wavelength and
scattering centre size gives the scattering angle distribution, the
\emph{phase function}. Rahman scattering, where the scattering centre is
affected by the scattered photon, and Brillouin scattering, where
photons are scattered on (thermal) density fluctuations, may result in a
change of photon energy. However, these processes are subdominant to Mie
scattering in both glacial ice~\cite{amandascat} and sea
water~\cite{antareswtr} where light is scattered by centres of
very different types and sizes: from ice crystal point defects to air
bubbles and mineral grains in ice, and from biological
matter to sediment particles in water.

In natural ice and water it is difficult to determine the phase
function from in situ measurements. Instead, the \amanda and
\textsc{Antares} collaborations have used calibration light sources to
determine the propagation characteristics assuming certain forms of the
scattering angle distributions. In the case of ice, a one parameter
Henyey-Greenstein (HG) phase function is often used, approximating Mie
scattering under the assumption that scattering is forward
peaked~\cite{hengreen}. For water, a two parameter phase function is
more useful. For this paper we mostly use the single parameter HG
phase function
\begin{equation}
\label{hgphase}
p_\mathrm{HG}(\cos\theta) =
\frac{1-\tau^2}{2(1+\tau^2-2\tau\cos{\theta})^\frac{3}{2}},
\end{equation}
which is completely characterised by the $\tau$ parameter, the
mean of the cosine of the scattering angle $\theta$,
\begin{equation}
 \tau \equiv \langle\cos\theta\rangle = \int p_\mathrm{HG}(\cos\theta) \cos\theta\, \de (\cos\theta).
\end{equation}

The \emph{absorption length}, $\lambda_{\abb}$,
and \emph{scattering length}, $\lambda_{\sct}$, are the mean free paths of
exponential distributions.
The probability density function for the path length $s$
to the next scatter is
\begin{equation}
\label{scatdist}
f_{\lambda_{\sct}}(s)\equiv
\frac{\de F(s)}{\de  s} =
\frac{ e^{\displaystyle -s/\lambda_{\sct}}}{\lambda_{\sct}},
\end{equation}
where $F(s)$ is the probability distribution function.

When determining ice or water optical properties, there is a degeneracy
between $\lambda_{\sct}$ and $\tau$. One therefore considers the \emph{effective
scattering length}, $\lambda_{\eff}$, defined as
\begin{equation}
\label{effscat}
\lambda_{\eff} = \frac{\lambda_{\sct}}{1-\tau},
\end{equation}
which in anisotropic scattering is analogous to the (geometric)
        scattering length $\lambda_{\sct}$ in isotropic scattering; it
        is the distance which light propagates through a turbid medium
        before the photon directions are completely randomised.
Consider a
        collimated light pulse injected into a non-absorbing medium.
In this case, the photons are on average scattered at successive steps of length
        $\lambda_{\sct}$
and the
        projection of the net velocity vector on the original
        direction is decreased on average by
        $\tau=\langle\cos\theta\rangle$ in each scattering step (in
        which all photons are scattered)~\cite{chandrasekhar}. Hence the
        injected light is effectively transported a forward distance of
\begin{equation}
 \lambda_{\sct} \sum_{i=0}^{\infty} \tau^i \rightarrow  \frac{\lambda_{\sct}}{1-\tau} = \lambda_{\eff},
\end{equation}
and $\lambda_{\eff}$ has a natural interpretation as the distance that
        the
centre of gravity of the photon cloud
advances, in the limit of many scatters.

\section{Monte Carlo simulation implementation}

\label{photomc}

In this section we present the main ingredients in our photon Monte
Carlo simulation implementation. The end product is a \emph{set} of
photon flux density tables describing the evolution of the light field
around a user-defined source. For a given light source, a user-specified
large number of photons is generated according to the source
characteristics. The photons are then tracked and their contribution to
the overall light field is determined and recorded in a cellular grid
throughout a user defined portion of space, the \emph{simulation volume}. The
sensor locations are not fixed, but are dynamically specified when
accessing the simulation results. The photon intensity
and time distributions are stored in a six dimensional binary table.
Four of these dimensions are for the spatial and temporal location in
the simulation volume with respect to the emission point. As the
acceptance of the light
sensors is assumed to be azimuthally symmetric,
around the vertical axis in heterogeneous media and around any axis in homogeneous media,
the
photon impact direction is characterised by the zenith angle alone,
constituting the fifth dimension. The sixth dimension is the angle from
the light source principal axis at which a photon is emitted (this
dimension can be used, for example, to reweight the flux tables for a
different emission profile). These latter two dimensions are usually
integrated over when the photon tables are used with detector
simulations.
In this case the wavelength and angular sensitivity of the detector
elements need to be folded in as the photons are recorded, using
recording weights
which can be specified in functional or tabular form. For each light
source position and orientation one table is produced. A \emph{set} of
tables describes a range of source locations and directions, valid for
the specified class of light sources and detectors.

\subsection{Media and light source properties}
\label{lightsec}

The parameters describing the  optical
medium ($\ngr$, $\nph$, $\tau$, $\lambda_{\eff}$, $\lambda_{\abb}$) are taken to be functions
of wavelength and a spatial dimension $Z$, typically specifying the ice
or water depth. Thus, the propagation medium is divided into horizontal
regions which differ by their optical properties. For media where the
single parameter HG approximation does not provide an adequate
description of the scattering it can be replaced by other
phase functions. An example of this is the treatment of sea
water in Section~\ref{watermodels}.

A single simulation run begins with the injection of a photon with
wavelength and emission direction chosen from user-specified probability
distributions and at a user-specified location. Our procedures support
point-like (infinitesimal) light sources, where all emitted photons
originate from the same point, and volume light sources where the
emission is distributed over a volume.
An example light source, essential for neutrino astrophysics, is that of
a \chv emitter. In \px{} a \chv emitter is a point-like
light source with a \chv wavelength spectrum and an angular emission in a \chv
cone~\cite{chref}. In this case the emission is azimuthally symmetric around the
principal axis of the \chv emitter. Closely related are light sources
composed of many short \chv emitting tracks, such as electromagnetic cascades.
Another category of point-like sources are laser or
LED light sources, and in Section~\ref{obscomparison} we compare our
simulation of such sources with observations. Continuous and extended
light sources are composed by integration of infinitesimal
sources. For example, the light distribution due to a relativistic muon
is composed by integrating a series of infinitesimal \chv emitters over
space and time, as described in Section~\ref{muonsection}.
Simulation results for both point-like and extended \chv emitters are presented in Section~\ref{neuastro}.

\subsection{Coordinate systems for photon flux recording}

The \chv light source example possesses cylindrical symmetry. Although
this symmetry is typically broken by the response of the propagation
medium and the detector, a cylindrical coordinate system aligned with
the source's symmetry axis is a natural choice for the recording
of the light flux, and it is therefore used in the following. In
addition to cylindrical ($\rho,l,\phi$) coordinates, we have also
included functionality to allow the flux to be recorded in spherical
($r,\vartheta,\phi$) or Cartesian ($\varepsilon_1,\varepsilon_2,\varepsilon_3$) coordinates. A grid over the
spatial coordinates defines cells in which the photon flux density and
time distributions are averaged. The user-specified region of space
covered by these cells constitutes the \emph{recording volume}.

\begin{figure*}[htb]
\center \includegraphics[width=0.50\textwidth]{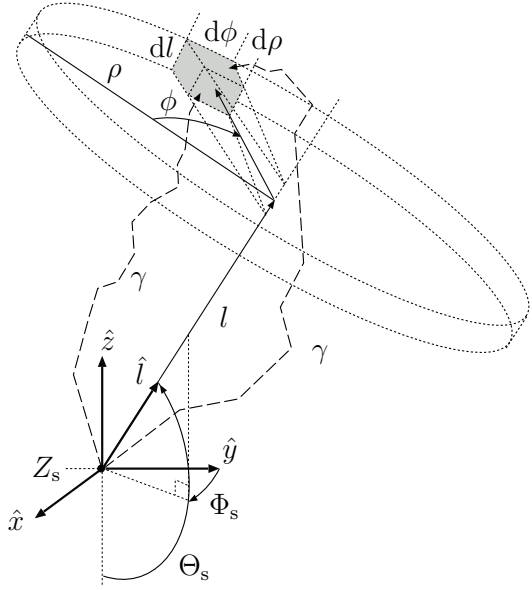}
\caption{ \label{unitcell}\label{binpicture} The recording cell geometry
with variables used for binning of photon flux data. Photons are emitted
from a user-defined light source at $Z_\src$ and their flux and time distributions
are recorded and averaged in all spatial cells the photons traverse (one of which is shown as
 a shaded volume).
These cells are defined in a coordinate system aligned with the source's
principle axis $\hat{l}$, which is tilted by $\pi-\Theta_{\src}$ with respect to the
medium symmetry axis $\hat{z}$. The angle $\phi$  is defined to be zero where
the radial vector is maximally aligned with $\hat{z}$.
 The azimuthal direction $\Phi_{\src}$ is degenerate since the medium
properties are assumed to be symmetric around the $\hat{z}$ axis.}

\end{figure*}

In some dimensions it can be desirable to have denser binning close to
the origin. Therefore, uniform as well as linearly increasing bin sizes
are supported for $\rho$, $l$, and the time $t$. In addition to
specifying the spatial coordinates where the light flux is recorded
relative to the source, the location of the source and the direction of
its axis of symmetry with respect to the medium are needed. These can be
characterised using just two coordinates, the depth $Z_{\src}$ and the
zenith angle $\Theta_{\src}$. This is because of the horizontal symmetry
of the medium, as discussed in Section~\ref{lightsec}.
Fig.~\ref{unitcell} shows the coordinates and a recording cell in which
the average flux is recorded. 

If in addition to the medium symmetry around $\hat{z}$ we assume that
the light source is symmetric with respect to the sign of $\phi$
(besides any light source that is azimuthally symmetric around its axis
this is the case for some non-isotropic LED calibration light sources
used in glacial ice \cite{amandaice,icecuberef}),
the flux needs only to be tabulated for $\phi$ between $0$\degr{} and
$180$\degr. For heterogeneous media we require that the sensor acceptance
and the medium properties are symmetric around the same axis $\hat{z}$.
This is not required for homogeneous media since the coordinate system can
be rotated 
%
%
%
make the sensor symmetry axis
collinear with 
the $\hat{z}$ axis used to define source location and orientation.
\px~can be extended to handle cases where
the two symmetry axes do not align in heterogeneous media, by adding further dimensions to the
photon tables.

\begin{table}[htb]
\begin{center}
\begin{tabular}{l c r r@{\usp}l r@{\usp}l}
\multicolumn{2}{c}{Dimension} & Bins & \multicolumn{2}{c}{Low} & \multicolumn{2}{c}{High}  \\
\hline
Radial & ${\rho}$ & 30  & 0&m   &  500&m      \\
Longitudinal & ${l}  $  & 51  & -500&m & 500&m      \\
Azimuthal & ${\phi}$ & 10  & 0&\degr    & 180&\degr      \\
Time & ${t} $ & 50 & 0&ns    & 6000&ns     \\
\end{tabular}
\caption{An example of table binning using cylindrical coordinates. The
rotational symmetry of the emitter and the horizontal symmetry of the
medium implies an azimuthal symmetry in $\phi$, so that the flux is the
same at $-\phi$ and $\phi$. \label{binexamplex}}
\end{center}
\end{table}

A binning example is shown in Table~\ref{binexamplex} for a cylindrical
coordinate system. As each single precision floating point number
requires four bytes, the size of this example table would be 3
megabytes. To get the total table set size this must be multiplied with
the number of light source positions $Z_{\src}$ and zenith angles
$\Theta_{\src}$ of interest. For example, with 50 source depths and 20
source angles, the total \emph{table set} size is 3 gigabytes.

\subsection{Photon propagation}

Starting at their point of origin, the photons are tracked and recorded
along straight-line paths between successive scattering points using one
of two methods. In the volume-density mode, photons are recorded
at equidistant recording points along their paths. 
In the area-crossing mode, photons are recorded at every
surface-crossing into a cell. In the latter mode, the propagation step is
dynamically calculated to bring the photon to its next cell boundary.
The photon's propagation properties are always updated at medium boundaries
and scattering points, regardless of recording method.

When a scattering point is reached, the photon's direction is changed by
an angle randomly drawn from the selected phase function, typically the
HG function in Eq.~(\ref{hgphase}), and in a uniform random azimuthal
direction. At this point, the distance to the next scattering point is
drawn from an exponentially distributed random variable with a mean
value of the local scattering length.

Photon absorption is taken into account by successively updating the
photon survival weight $w$ as the photons propagate through regions of
different absorption. For a photon tracked $n$ steps, each
within a locally homogeneous medium, the weight is given by
\begin{equation}
\label{absweights}
w = \prod^n_{i=0} \exp{\left(-\frac{\Delta s_i}{\lambda_{\abb,i}}\right)},
\end{equation}
where $\Delta s_i$ is the length of step $i$ in a region with absorption length $\lambda_{\abb,i}$.
The weight is updated every time the photon is scattered, enters a new
medium region, or is recorded.

When a photon enters a medium region with different scattering and
absorption parameters, these are updated, from
($\lambda_{\sct},\lambda_\abb$) to ($\lambda'_{\sct},\lambda'_\abb$), at
the region boundary. At this point, the remaining distance to the next
scatter is $s'=s \lambda_{\sct}'/\lambda_{\sct}$, where $s$ would have
been the remaining distance to the next scatter in the former region
with scattering length $\lambda_{\sct}$. Refraction at the boundary is
supported but reflection is ignored since it is assumed that refractive
index variations are continuous.

\label{fluxsec}

During its propagation, the flux contributed by a photon is recorded either in
each spatial cell it enters (in area-cossing mode) or each time it completes
a propagation step (in volume-density mode).
The photon flux $\Phi$ (particles per area
and time) at any point $(\rho,l,\phi)$ at a time $T$ after the emission
from a light source at a depth $Z_{\src}$ pointing in a direction
$\Theta_{\src}$ is denoted $\Phi(Z_{\src},\Theta_{\src},\rho, l,
\phi,T-t_0(\rho, l))$, where $t_0$ is a reference time typically chosen
to be the first time causally connected to the light emitted by the
source,
\begin{equation}\label{residualtime}
 t_0(\rho,l) = \frac{n_{\mathrm{ref}} }{c }\sqrt{\rho^2 + l^2},
\end{equation}
where $c$ is the speed of light in vacuum and $n_{\mathrm{ref}}$ is a
user-specified reference refractive index. This reference time convention
is appropriate for point-like stationary light sources only.
For fast moving light sources such as muons a different expression is
used which takes into account the more complicated causality condition.
The {\it residual time} $t\equiv T-t_0$ is the time delay caused by
scattering, relative to the propagation time for a photon travelling in a
straight line.
Photons are tracked until their residual times exceed a user-specified
value.
The tracking of a photon is terminated if it leaves the simulation volume
(which can be larger than the recording volume to allow the photon to
scatter back into the recording volume) or if its survival weight drops
below a pre-set value.

The probability density function for a photon flux at time $t$
is given by
\begin{equation}\label{pdfdef}
f_{\ppdf}(Z_{\src},\Theta_{\src},\rho, l, \phi,t)=\frac{\Phi(Z_{\src},\Theta_{\src},\rho, l, \phi,t)}%
{I(Z_{\src},\Theta_{\src},\rho, l, \phi)},
\end{equation}
where $I$ is the time integrated photon flux, or intensity,
\begin{equation}
  I(Z_{\src},\Theta_{\src},\rho, l, \phi)=\int_{-\infty}^{\infty} \Phi(Z_{\src},\Theta_{\src},\rho, l, \phi,t) dt.
\end{equation}

Since photon fluxes are additive we can determine the time
distribution of a combination of light sources through
\begin{equation}
\label{combopdf}
 f_{\ppdf}(t) = \frac{\sum_i I_i f_i(t) }{\sum_i I_i} = \frac{\sum_i \Phi_i (t) }{\sum_i I_i}.
\end{equation}

The way in which the photon intensity in a spatial cell is calculated
depends on the recording method. In the area-crossing method, the
contribution of each photon to the total flux in the cell depends on the
projected surface area of the cell as seen in the direction of the
contributing photon as it crosses the cell boundary. In the
volume-density method, photons contribute at equidistant recording
points along their paths, so that the contribution is proportional to
the number of recording points that fall in a given cell. The respective
equations for the calculation of the observed intensity, per emitted
photon, are
\begin{eqnarray}
I_{\scriptscriptstyle \mathrm{Area Crossing}}&=&\frac{1}{N} \sum_{\gamma=1}^N w_\gamma A_\bot^{-1}, \\
I_{\scriptscriptstyle \mathrm{Volume Density}}&=&\frac{1}{N}
 \sum_{\gamma=1}^N \sum_{k=1}^{n_\gamma} w_{\gamma}(k) \frac{\Delta s }{V},
\end{eqnarray}
where the $\gamma$ sum runs over the $N$ simulated photons, $A_\bot$ is
the recording cell area projected perpendicularly to the direction of
the photon, $V$ is the volume of the cell, and $\Delta s$ is the
recording point separation along the photon path in the volume-density
method. The quantity $n_\gamma$ is the number of recording points inside the
cell for a given photon. 
Hence, in the area-crossing mode, photons are recorded at every
surface-crossing into a cell, whereas in the volume-density mode, they
are recorded at equidistant recording points. 
The photon weight $w_\gamma$ is a product of
the absorption-induced survival weight, Eq.~(\ref{absweights}), and
optional user-specified sensor detection efficiencies.

The calculation and recording of a photon's contribution to the flux in
a recording cell is computationally expensive, but a suitable choice of
method (area-crossing or volume-density) can speed up the simulation. To
optimise for speed one compares the step size $\Delta s$ with the scale
dimension of a recording cell $D_\mathrm{cell}$. If 
$\Delta s < D_\mathrm{cell}$, the area-crossing method is competitive; 
otherwise 
the volume-density method should be used. Hence the area-crossing method
can result in faster code execution for large detection volumes with
large recording cells, while the volume-density method is faster for
small, dense recording cell configurations. To ensure unbiased sampling
in volume-density mode even for $\Delta s \gg D_\mathrm{cell}$, the path
length to the first recording point after emission is drawn from a
uniform distribution between 0 and $\Delta s$. 
To maximise the number of independent sampling points for a given
execution time, $\Delta s$ is balanced against the total number of photons
$N$. A large $\Delta s$ allows a large $N$, but if too large the number
of recordings per execution time will drop as more time is spent
propagating photons between recording points. Using $\Delta s \approx
\lambda_{\sct}$ is often a good compromise.

\label{verify}
Another way to improve the speed of the Monte Carlo simulation
is through importance-weighted scattering. This means that the photons
are propagated using scattering parameters which are different from
those of the scattering situation at hand in order to get higher
statistics at
low probability phase space locations. When solving a Monte Carlo
problem for random numbers $x_i$ (for example, scattering angles) from a
probability density distribution $f_1(x)$, we can choose to instead
sample from another distribution $f_2(x)$ while applying a weight of
$f_1(x_i)/f_2(x_i)$.
As an example, a user might want to oversample straighter paths to
enhance the statistics for early photons at large source-receiver
distances. The user could then simply scale the scattering length by a
factor $k > 1$, propagate photons with scattering described by
$f_{k\lambda_{\sct}}(s)$ (see Eq.~(\ref{scatdist})), and reweight each photon by
$f_{\lambda_{\sct}}(s)/f_{k\lambda_{\sct}}(s)$ for every scatter it experiences.
Alternatively, the mean cosine of the scattering angle, $\tau$, can be
modified to achieve a similar effect using Eq.~(\ref{effscat}). The
effective scattering length is made a factor $k$ longer by $\tau' =
(k-1+\tau)/k$. The corresponding weight is
$f_{\tau}(\theta)/f_{\tau'}(\theta)$.

We have described how the photon flux density is calculated in cells
throughout the simulation volume. In neutrino astronomy, the quantity of
interest is the number of photons detected by an optical sensor such
as a photomultiplier. Since the sensor response is usually wavelength and angle
dependent but we do not want to store the wavelengths and arrival
directions of the simulated photons, \px{} includes the option to fold
the sensor response into the simulation when the tables are generated. When
this option is selected, user-supplied wavelength and angular efficiency
files are used to weight the photon flux density to obtain the detected
photon flux.

\subsection{Propagating light sources}
\label{muonsection}

To this point we have discussed how to obtain a set of tables
describing the photon fluxes for a range of locations and orientations
of a point-like, stationary light source.
A propagating light source can not be satisfactorily
approximated as a flash of light from a single spatial point.
We discuss in the following the modelling of \chv light from high energy muons,
which give rise to kilometre
scale \chv emitting tracks in water or ice. Our method,
however, applies also to other line-like light sources such as high
energy tauons.

To calculate the photon flux distribution generated by a muon we
integrate over the photon flux distributions of many point-like \chv
emitters with $\Theta_{\src}$ given by the muon direction.
 A set of point-like photon tables provide
the differential light flux $\Phi_{\mathrm{point}}$ at any space-time
location from sources at any causally connected location. This serves
as integration kernel,
\begin{equation}
G_{\mathrm{point}}(\vect{x}_{\src}(t_{\src}),\:\vect{x}_{\rec},t_\rec)
 \equiv \frac{\partial}{\partial t_{\src}} \Phi_\mathrm{point}(\vect{x}_{\src}(t_{\src}),\:\vect{x}_{\rec},t_\rec),
\end{equation}
 where ($\vect{x}_{\src},t_{\src}$) and
($\vect{x}_{\rec},t_\rec$) are the emission and receiver coordinates. The
light
distribution of a propagating muon is thus generated by convolving
this kernel with the track of the muon, $\vect{x}_\mu(t_{\mu})$,
so that
\begin{equation}
\label{finitesum}
 \Phi_{\mu}(\vect{x}_\rec,t_\rec) =
  \int_{t_{\mu_\mathrm{start}}}^{t_{\mu_\mathrm{stop}}}
\!\! G_{\mathrm{point}}(\vect{x}_{\mu}(t_{\mu}),\:\vect{x}_{\rec},t_{\rec}) \: dt_{\mu}.
\end{equation}
\px{} provides the capability to efficiently perform this integration for a series of fixed muon
directions and locations and to store the resulting light fluxes in tables like those
for the point-like light sources.
These tables are then used to obtain the light flux for muon tracks through any part
of the simulation volume by interpolation (Section~\ref{photolib}).
Although the construction of muon events through the integration in Eq.~(\ref{finitesum}) can
in principle also be done event by event in a detector simulation or event reconstruction, such
an approach would often be significantly slower since the number of considered
events, and therefore the number of required flux calculations, is
typically larger than the number of fixed muon directions and locations needed to adequately cover
the relevant range.

\begin{figure*}[htb]
\center
\includegraphics[width=\textwidth]{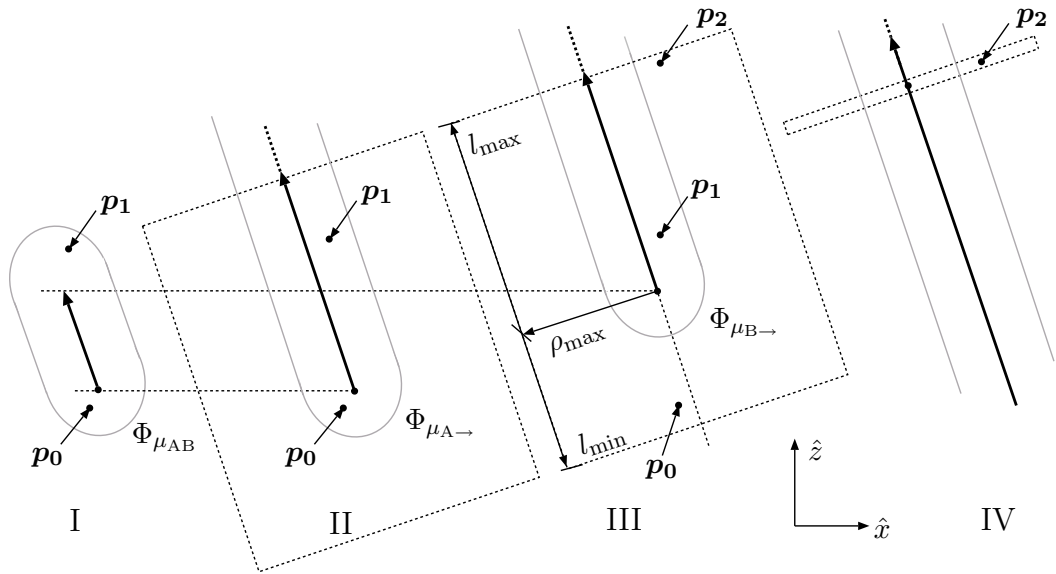}
\caption{
\label{muonsubtractpic}
Schematic view of flux tables for four collinear tracks (offset in $x$
 for clarity).
The light
distribution of a finite track (I) is the
difference between  two semi-infinite tracks (II) and (III). The dotted
rectangles show the outer limits of the semi-infinite tables.
The gray lines represent example isointensity contours. The representation of photon flux
at a point $\vect{p}_2$ far from the starting and
stopping points, can be done in two ways. Either, as in (III), by
considering the edge of the best matching semi-infinite table, or, as in (IV), with an
\emph{infinite} track table with only a single $l$ bin.}
\end{figure*}

To provide the photon flux for any given track length
without having to dynamically perform the time consuming integration of
Eq.~(\ref{finitesum}), we have developed a scheme based on the
subtraction of semi-infinite tracks.
The flux at ($\vect{x}_\rec,t_\rec$) arising from a finite muon starting
at $\vect{x}_{\iA}$ and stopping at $\vect{x}_{\iB}$ can be expressed as
the difference between two
semi-infinite tracks. For two
semi-infinite \emph{starting} tracks, one (denoted $\mu_{\iA\rightarrow}$) starting
at the point $\vect{x}_{\iA}$ and the other ($\mu_{\iB\rightarrow}$) starting
at $\vect{x}_{\iB}$ located further along the muon track, we write
\begin{equation}
\label{difftrk}
\label{muonpdf}
 \Phi_{\mu_{\iAB}}(\vect{x}_\rec,t_\rec) =
 \Phi_{\mu_{\iA\rightarrow}}(\vect{x}_\rec,t_\rec) -
 \Phi_{\mu_{\iB\rightarrow}}(\vect{x}_\rec,t_\rec),
\end{equation} and the probability density function can be written
\begin{equation}
 f_{\mu_{\iAB}} = \frac{f_{\mu_{\iA\rightarrow}}-f_{\mu_{\iB\rightarrow}}}{
I_{\mu_{\iA\rightarrow}}-I_{\mu_{\iB\rightarrow}}}
\end{equation}
using Eq.~(\ref{pdfdef}).
This construction of finite tracks is depicted in Fig.~\ref{muonsubtractpic}. %
Consider the point $\vect{p}_0$, close to the finite
track start point.
At this point, the contribution from
$\Phi_{\mu_{\iB\rightarrow}}$ is comparably small, so that
$\Phi_{\mu_{\iAB}} \approx \Phi_{\mu_{\iA\rightarrow}}$. However,
for the point $\vect{p}_1$ close to the end of the finite track,
$\Phi_{\mu_{\iA\rightarrow}} \approx \Phi_{\mu_{\iB\rightarrow}}$ and the calculation of $\Phi_{\mu_{\iAB}}$ becomes sensitive to exact
numerical cancellation. It is then
numerically superior to subtract two semi-infinite
\emph{stopping} tracks, $\mu_{\rightarrow\iA}$ and
$\mu_{\rightarrow\iB}$,
and write $ \Phi_{\mu_{\iAB}} =
\Phi_{\mu_{\rightarrow\iB}}-\Phi_{\mu_{\rightarrow\iA}}$.
Our algorithm dynamically chooses between these two descriptions
depending on which of the endpoints is closer to the
observation point.

For applications in which tracks can be regarded as completely infinite,
the flux tables are made smaller by removing redundant information.
At a given observation point in the medium, the flux from an infinite
muon track with table origin $Z_1$, giving the point a longitudinal
coordinate $l_1$, is identical to the flux in a table with origin $Z_2$
where the point is at \mbox{$l_2 = l_1+(Z_2-Z_1)/\cos\Theta_{\src}$}.
Therefore in each infinite muon table only one $l$ bin is retained,
typically at $l=0$. 
Fig.~\ref{muonsubtractpic} illustrates both ways of accessing
information about virtually infinite tracks: with infinite tables and
with semi-infinite tables.

The size of a set of infinite tables is typically about 1\% of the size of the
corresponding semi-infinite description, which can be
tens of gigabytes.
Hence the infinite tables can easily be loaded into computer memory all
at once, which is particularly useful in event reconstruction where it
can be hard to estimate in advance the properties of an event to be
reconstructed.
Reconstruction and large table
support is discussed in the following section, describing the \px{}
reader library.

\label{infmuons}

\subsection{Using the photon flux tables for event simulation and reconstruction}

\label{photolib}

For event simulation and reconstruction, the photon flux tables are
accessed using either a set of \mbox{\textsc{ansi-c}} procedures, or a
more abstract (\textsc{root} compliant) user interface written in \cpp,
both provided with the \px{} package.

The full simulation of for example an ultra relativistic muon crossing
the detector volume is performed by first propagating the muon through
the detector medium with a
charged particle propagator such as {MMC}~\cite{mmc}. This results in a
list of light generating subevents such as minimum ionising muon track segments and
associated electromagnetic showers induced by stochastic energy losses.
The detector specific simulation program can then query the
corresponding \px{} table information to obtain the number and time
distribution of detected photons in each detector module from each such
subevent. \px{} comes with a variety of idealised light emission
profiles, such as those of minimum ionising muons and point-like
electromagnetic and hadronic showers.

A set of tables is loaded into memory using the table reader library.
The user can then query the photon flux tables by specifying the location
($\vect{x}_{\src}$) and orientation ($\Theta_{\src},\Phi_{\src}$) of the
source, and the location of the light detectors ($\vect{x}_{\rec}$). Two
additional source characteristics are optional in the query:
the source length $L$ (applicable to finite muons) and the
energy $E$ used to scale the light source intensity.
In an experiment simulation, the user first requests the expected number
of detected photons $N(\vect{x}_{\src},\Theta_{\src},\Phi_{\src},\vect{x}_{\rec},{E},{L})$.
This query also returns a table
reference which can be used to get photon arrival times randomly drawn
from the tabulated time distribution at the corresponding coordinates.
For event reconstruction, \px{} also provides the arrival time
probability density function
$f_{\ppdf}(\vect{x}_{\src},\Theta_{\src},\Phi_{\src},\vect{x}_{\rec},L,t)$, which
can be used by track-fitting algorithms, for example maximum-likelihood
routines. Both $f_{\ppdf}$ and the photon intensity (giving $N$) are naturally continuous
in $L$ and $E$, and are made continuous in
$\vect{x}_{\src},\Theta_{\src},\Phi_{\src},\vect{x}_{\rec}$, and $t$ by multidimensional
linear interpolation. The interpolation of time distributions is flux
weighted, in agreement with Eq.~(\ref{combopdf}),
\begin{equation}
f_{\ppdf} = \frac{\sum_i \omega_i I_i f_i}{\sum_i \omega_i I_i}
\textrm{, with~} \textstyle \sum_i \omega_i =1,
\end{equation}
where $\omega_i$ are the interpolation weights. 
When the flux for a requested source direction $\Theta_{\src}$ is
interpolated from two surrounding tabulated directions $\Theta_{1}$ and
$\Theta_{2}$, these tables are first (implicitly) rotated to $\Theta_{\src}$.
The receiver coordinates $\vect{x}_{\rec}$ are then identical for the
$\Theta_{1}$ and $\Theta_{2}$ table. An analogous approach is used for the
source origin $\vect{x}_{\src}$.

It is sometimes necessary to convolve the photon time distributions with
the detector time response function or the emission time profile. This
can be done with a provided routine operating on the photon flux tables.
Convolution with a Gaussian or with one of two light source time
distributions with longer positive tails \cite{antareswtr} (see
Section~\ref{obscomparison}) has been implemented.

Any number of table sets can be loaded simultaneously (limited only by
available memory), making it possible to simulate or reconstruct the
cumulative signal from different types of light sources (and detectors).
This is useful for example when simulating muons with secondary
electromagnetic showers from a primary muon track or several coincident
muons. Since detailed photon table sets are often many times larger than
the primary memory of a computer, we provide some memory management
tools. Users can dynamically load and unload tables, and select loading
of tables corresponding to limited ranges of depths ($Z_{\src}$) and
light source angles ($\Theta_{\src}$). In experiment simulations this is
particularly useful since the parameters for the simulated particles are
known. In event reconstruction, the loaded tables should cover the phase
space taken into account in the fitting algorithms. If memory
limitations restrict this phase space, reconstruction can be performed
for subregions, on events predetermined to lie within subregions small
enough for the corresponding photon tables to fit in memory. Such
preselection can be done with a set of more coarsely binned \px{} tables
or with other first-guess approaches (for muons, the much smaller tables
of the \emph{infinite} description can be used, see
Section~\ref{infmuons}).

In addition to what has already been mentioned, the \px{} package includes
several other tools for processing of the photon flux tables,
including integration in any dimension and conversion between
differential and cumulative time distributions.

\section{Comparison with observations}
\label{obscomparison} In this section we use measurements with
artificial calibration light sources in sea water and glacial ice to demonstrate that \px{}
reproduces the general behaviour of the observed photon
distributions.

\subsection{Modelling of natural water and {\sc Antares}' Mediterranean water surveys}
\label{watermodels}

The {\sc Antares} collaboration is constructing a 0.1 km$^2$ water
neutrino detector in the deep Mediterranean sea ~\cite{antareswtr}. In
the present design the detector has 12 vertical strings, each of which
has an instrumented height of about 350~m and consists of 25 storeys
with three optical sensors each.
Three strings were deployed in 2006 and the remaining strings are
scheduled to be installed during 2007.

\begin{figure*}[htb]
\includegraphics[width=\textwidth]{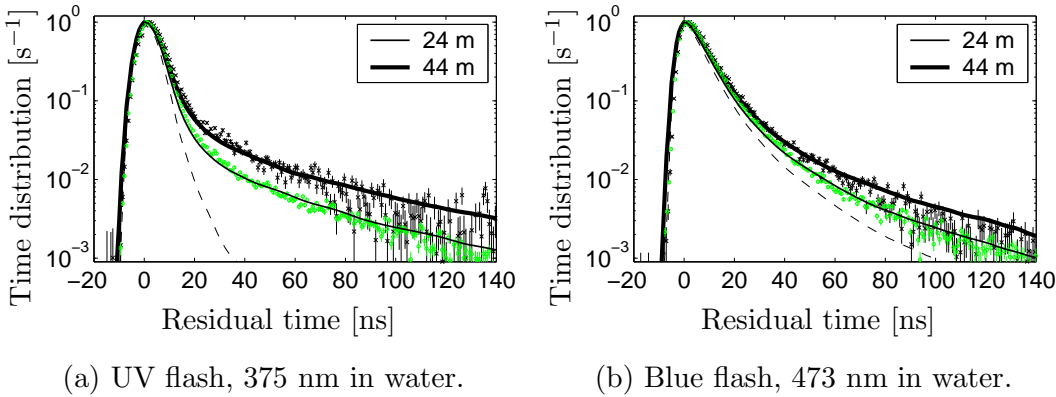}
\caption{
\label{waterpicture} Residual time distributions at two distances
from
monochromatic flashes in water. The circles and stars are calibration
data from ANTARES~\cite{antareswtr}, while the solid curves are our simulation
results using the measured water properties.
The emission time profiles of the light sources (dashed curves) were measured
in air, where scattering and absorption can be ignored. The
distributions are normalised to unity at the peak value.}
\end{figure*}

Absorption and effective scattering lengths in the deep Mediterranean
sea water have been investigated by the {\sc Antares} collaboration~\cite{antareswtr}. The surveys were performed during
several seasons using a calibrated setup of isotropic light sources, one
in the blue at 473 nm and one in the UV at 375 nm. For blue (UV)
a $\lambda_{\abb}$ of 60 (26) m and a $\lambda_{\eff}$ of 265 (122) m with
15\% time variability are quoted by {\sc Antares}. The details of the experimental setup, such as the
light emission time profile of the source and detector
efficiencies, play a large role for the photon flux
time distribution profile in media, like ocean water, where scattering
is weaker than absorption. In addition, since the light is typically
observed at distances shorter than or comparable to the effective
scattering length, the scattering phase function plays an even larger
role. {\sc Antares} assumed a weighted sum of a Petzold and a molecular
(Einstein-Smoluchowski) distribution. The Petzold distribution has
$\tau=0.924$, and is here approximated by a HG distribution, while the
molecular distribution is approximated by isotropic scattering (HG with
$\tau=0$). Fig.~\ref{waterpicture} compares our simulation results
using this model
with \textsc{Antares} measurements from June
2000~\cite{antareswtr}. The agreement verifies the validity of the model
 and our photon flux simulation for this case.

\subsection{Modelling of ice and {\sc IceCube}'s  Antarctic glacial ice surveys}

\label{icemodels}

The IceCube neutrino telescope, under construction deep in the
glacial ice at the geographic South Pole, is planned to become a
high-energy neutrino detector of 1 km$^3$ instrumented
volume~\cite{icecuberef}. It is planned to have 80 strings, of which 22
have been deployed as of January 2007, each equipped with 60 encapsulated photomultipliers
evenly distributed over depths between 1450~m and 2450~m. IceCube builds on to the 19
strings of the \amanda array, which have been in operation since 2000.

A detailed study of the properties of the deep South Pole glacial ice
has been performed by the \amanda collaboration~\cite{amandaice}. The
glacial ice is very clear in the optical and near UV with absorption
lengths of 20--120~m depending on wavelength. At wavelengths shorter
than $\sim\!210$~nm and longer than $\sim\!500$~nm, absorption is
dominated by the properties of pure ice, while in the intermediate range
absorption by impurities dominates. The effective scattering length is
on average 25~m, less for shorter wavelengths. Both scattering and
absorption are strongly depth dependent and vary on all depth scales.
The variations at depths exceeding 1450~m, where bubbles no longer
exist, are explained by varying concentrations of insoluble dust
deposits which correlate with changes in climatic conditions over
geological time scales. By using physics motivated functional forms for
the wavelength and depth dependences of scattering and absorption, the
\amanda collaboration has elaborated a heterogeneous ice model~\cite{amandaice}
by investigating a large number of recorded light
distributions generated by in situ pulsed and steady light sources at
different wavelengths.
The resulting effective scattering and absorptions lengths,
$\lambda_{\eff}$ and $\lambda_{\abb}$ as functions of wavelength, were
averaged in 10~m depth intervals.

\begin{figure*}[htb]
\includegraphics[width=\textwidth]{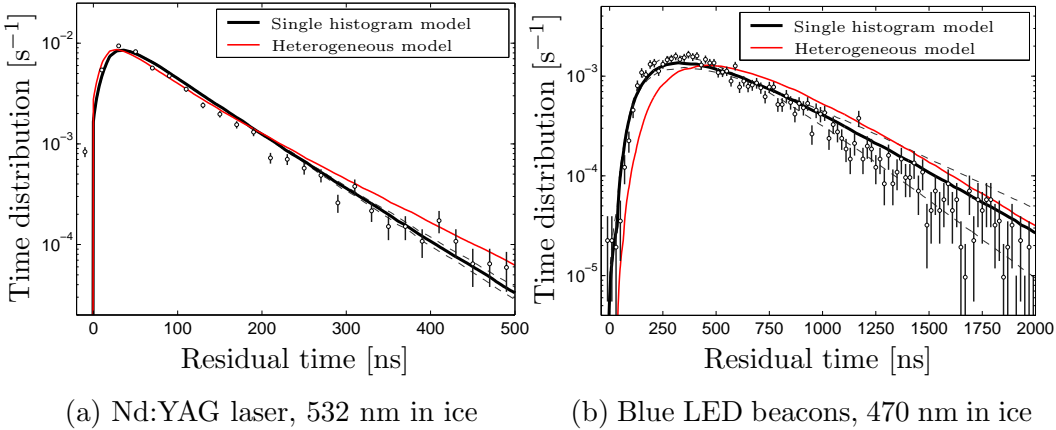}
\caption{
\label{iceplots}Residual time distributions for two monochromatic pulsed light sources in
deep glacial ice.
In (a), light is recorded at a horizontal distance of 75~m from an
isotropic laser source located at a depth of 1825~m.
In (b), the detector is at a horizontal distance of 140~m from an
upward-pointing LED emitter located at a depth of 1580~m.
The \amanda calibration data are shown with Poissonian error bars.
The intrinsic timing widths of the light sources, less than 10~ns, were not included in
the simulation.
The thick solid curves are \px{} simulations using the scattering and absorption parameters fitted to
these particular time distributions, and the thin dashed curves
represent two opposite parameter variations within the parameter uncertainties from the fits.
The thin solid curves show our simulation results with the heterogeneous ice model~\cite{amandaice}.
}
\end{figure*}

Using the \amanda ice model parameters, we have generated simulated time
distributions corresponding to two combinations of wavelength and
light source--receiver positions, and
compare them with experimental distributions in Fig.~\ref{iceplots}.
The thick solid curves are our results when using the scattering and
absorption parameters fitted to these particular delay time distributions~\cite{amandaice},
with dashed curves representing two
opposing deviations within the parameter uncertainty from these fits.
The parameters fitted to the displayed distributions were
$\lambda_{\eff}=27.6$~m, $\lambda_{\abb}=20.5$~m for the 532~nm curve, and
$\lambda_{\eff}=22.6$~m, $\lambda_{\abb}=82.0$~m for the 470~nm
curve, both with $\tau=0.94$. The slight overestimation in the tail in
Fig.~\ref{iceplots}(b) is due to systematic uncertainties in the simulation of the LED emitter
which lead to an imperfect description of the data in this particular case.
The thin solid curves are
simulation results with the heterogeneous ice model which is based on
data from many source--receiver combinations at different depths and
wavelengths. The differences between
the two solid curves in each picture reflect the fact that the ice model parameters
are averages over all fitted parameters in 10~m depth bins and the parameters from individual fits
are distributed around these averages.

\section{Application to neutrino astronomy}
\label{neuastro}
\label{glacialphysics}

In neutrino astronomy, the universe is studied using high energy
neutrinos as cosmic messengers. The neutrinos can be detected optically only after
interacting with matter in the vicinity of the neutrino telescope and
producing charged, \chv light emitting particles like muons. Some of the
emitted light is recorded by optical sensors distributed throughout the
detector volume.

Ultra-relativistic muons are the primary channel through which high
energy neutrinos are detected by optical neutrino telescopes. They are
also the primary background in the form of so-called atmospheric muons
arising from high energy cosmic-ray interactions in the Earth's
atmosphere. An extraterrestrial neutrino signal is distinguished from the
background of neutrinos and muons created in the atmosphere mainly by
differences in energy spectra and angular distributions. It is therefore
important to establish the particle energy and direction in every
recorded event as accurately as possible. Our software contributes to
this aim by providing means for detailed photon flux simulations, using
depth and wavelength dependent optical properties as established at a specific site.
      High-energy neutrino telescopes are typically recording data
       for several years while they are constructed by
       adding more optical sensors. Photon flux tables generated by
       \px{} can cover arbitrarily large volumes and their use
       is therefore easily scalable to such growing sensor arrays.

To further
illustrate the utility of \px{}, we present in this section photon
propagation results for the inhomogeneous ice at the site of the IceCube
neutrino telescope~\cite{amandaice}.
The flux in all figures is given per emitted photon, and is weighted
with the angular and wavelength dependent acceptance of the optical
detectors used by IceCube, so that the figures display the expected number of detections
normalised to a 1~m$^2$ detection area in the direction of maximum optical
module sensitivity.
The photon flux tables were also convolved with a
10~ns wide Gaussian to account for photomultiplier jitter.

\begin{figure*}[htb]
\center
\includegraphics[width=\textwidth]{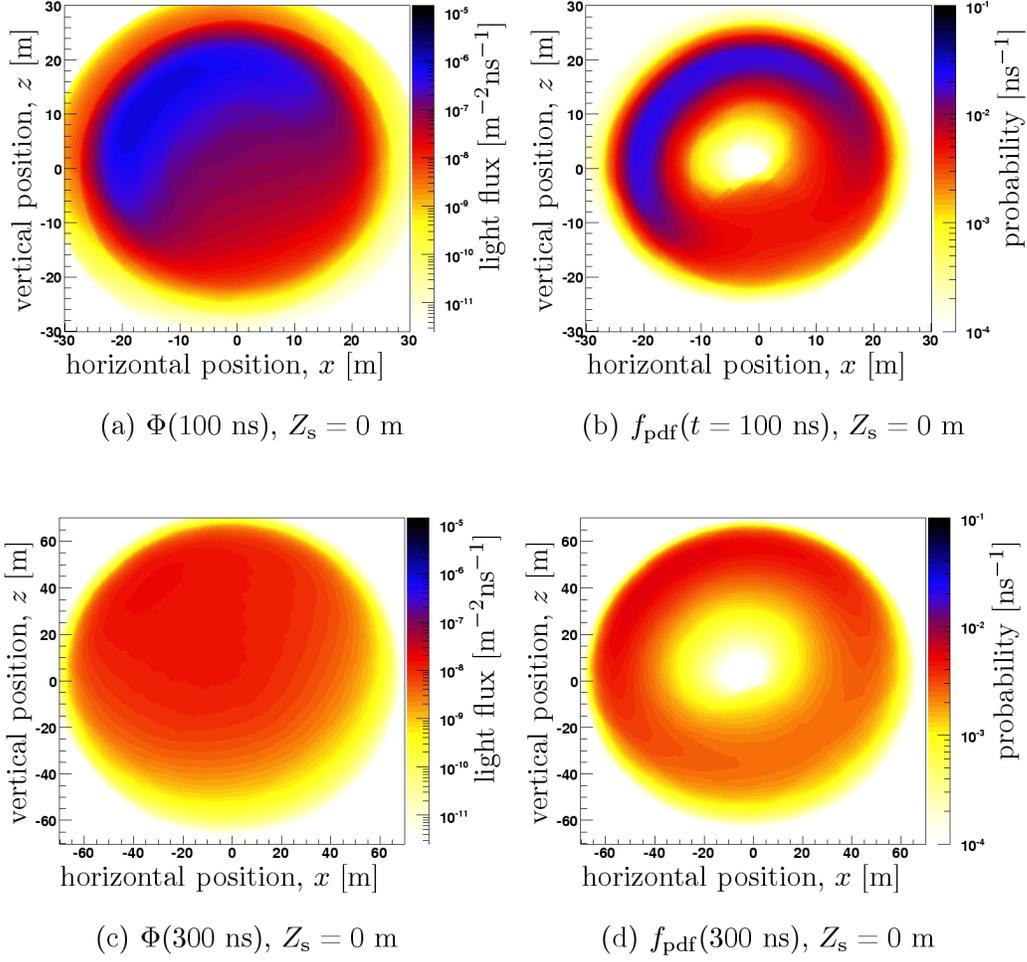}
\caption{\label{a350tx}\label{cascades}
Light flux generated by a simulated electromagnetic cascade near the centre of the
\amanda telescope.
The upper panel shows a vertical slice of the photon flux (left) and the
probability distribution (right) at $t=100$~ns after light emission from a
cascade at the origin and oriented toward the upper left at $\Theta_{\src}= 135 \degr$.
The lower panel shows the same distributions at $t=300$~ns.
Note the different scales in the two panels.
}
\end{figure*}

Fig.~\ref{cascades} shows the photon flux from a simulated infinitesimal
electromagnetic cascade at $1730$~m depth ($Z_{\src} = 0$ in \amanda detector coordinates).
Cascades are initiated in muon energy loss
processes as well as in primary neutrino interactions. The light spectra
of hadronic and electromagnetic cascades are \chv in nature, but
the light originates from many \chv light emitting particles. The \chv{}
 emission cone is slightly distorted~\cite{CHWthesis} since not all emitting particles
travel in parallel or at the same speed.
The cascade in Fig.~\ref{cascades} is oriented toward the surface, at $\Theta_{\src}=135$\degr,
pointing at the upper left corner of the picture. A vertical slice through the photon flux containing the
principal axis of the light source is displayed. The angular
distribution of emitted photons is peaked at the \chv angle, but the
photon flux is smoothed out by scattering as it evolves through the ice.
After 100 ns, we can still observe a peaked light distribution in the
forward direction. At later times, the flux becomes more and more
isotropic, to asymptotically resemble that of an isotropic flash.

Since the IceCube sensors are pointing downward they detect
upgoing light with a higher efficiency than downgoing light, which
needs to be scattered to reach the photomultiplier photocathode. As a result,
a point-like light source appears more upward.
This can be seen in Fig.~\ref{cascades} where the
direction of the light source appears to be oriented at an angle
larger than $\Theta_{\src}= 135 \degr$.

\begin{figure*}[htb]
\includegraphics[width=\textwidth]{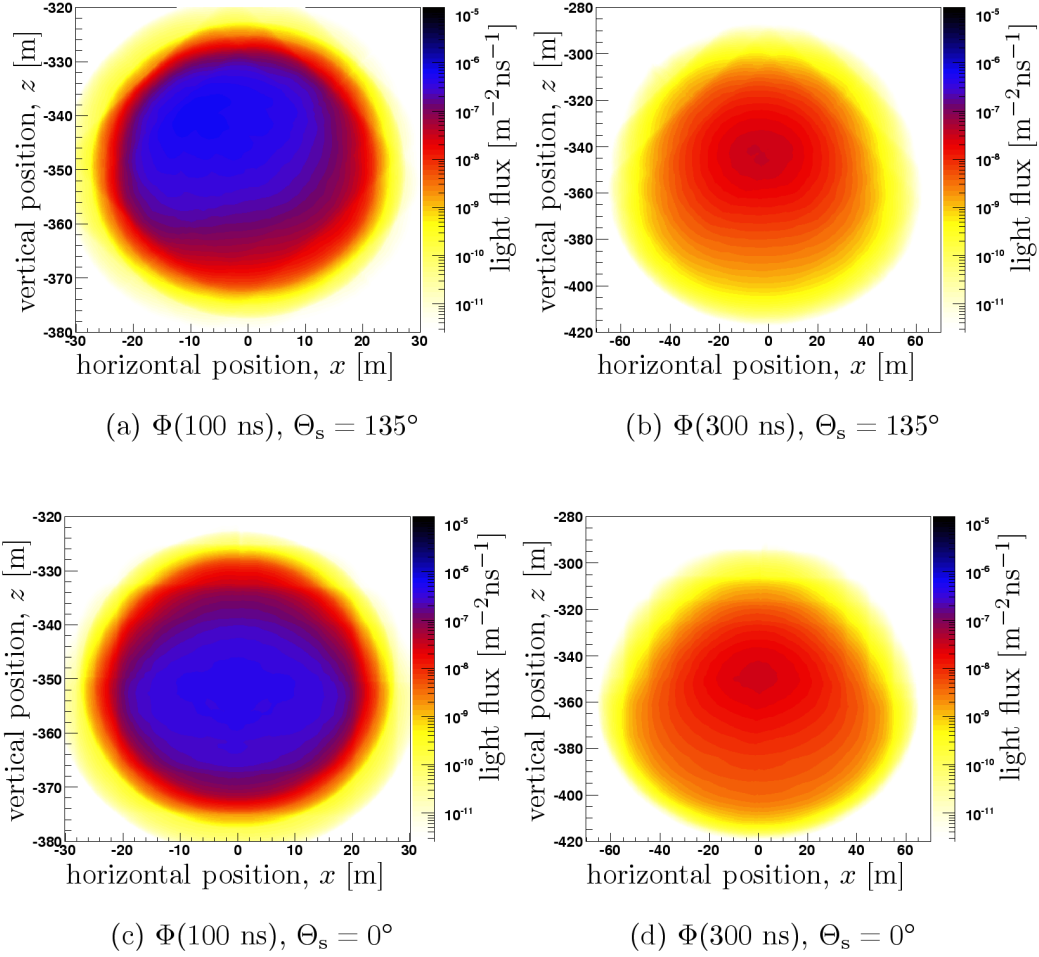}
\caption{
Light fluxes generated by two simulated electromagnetic cascades
$350$~m below the \amanda centre.
This location is immediately below a region with higher dust concentration
that causes stronger scattering and absorption.
The flux is shown for both cascades at 100~ns and 300~ns after light emission.
Because of the strong scattering
and absorption just above the emission point, the flux from the upward
pointing ($\Theta_{\src}=135$\degr) source in the upper panel, (a) and (b),
evolves from a distribution peaked at the \chv angle to a distribution similar to that from
a downward pointing ($\Theta_{\src}=0$\degr) source, shown in the lower panel, (c) and (d).
\label{cascadeslow}}
\end{figure*}

While the ice is very clear in the centre of the \amanda telescope,
there are other depths where stronger scattering and absorption distort
the photon flux. In the upper panel of Fig.~\ref{cascadeslow} simulation results
are shown for a setup analogous to that of Fig.~\ref{cascades} but
with the shower origin 350~m deeper in the ice. This location is
immediately below a region of strong scattering and absorption. The
cascade direction is again $\Theta_{\src}=135$\degr, but because of the particular
medium properties in this region the event shape appears to be more isotropic and even
resembles a downward pointing cascade. However, it does differ from truly downward pointing
events at the same location, shown in the lower panel of Fig.~\ref{cascadeslow}.
In this particular case it is exceptionally hard to characterise the
event correctly, but by using the correct heterogeneous medium description
we improve the possibility to distinguish these cases. This is important
for event simulation and reconstruction of parameters such as the zenith
angle and the cascade energy.

\begin{figure*}[htb]
\center
\includegraphics[width=\textwidth]{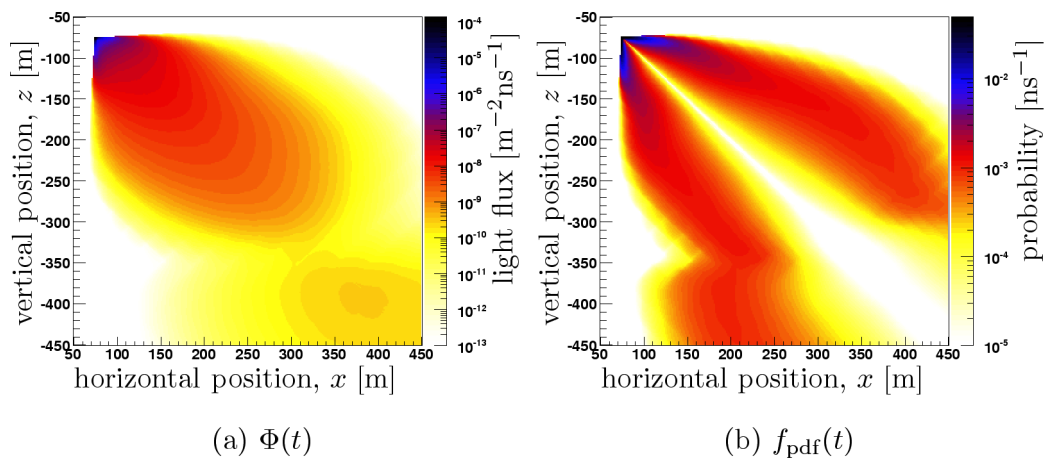}
\caption{
\label{infmuonpic} A snapshot of the light
field produced by a muon which entered from below, at
an angle $\Theta_{\src}=135$\degr.
Inhomogeneities in the medium properties distort the smoothly arched
light cone, as is most easily seen in the probability density function (b) in the particularly dusty region
around $z=-350$~m which has stronger scattering and absorption.
In (a), the flux is depleted in the dusty region.
}
\end{figure*}

Inhomogeneities in the detector medium also strongly affect the optical topology
of muon events.
Fig.~\ref{infmuonpic} shows the light
distribution of a simulated muon moving upward through deep South Pole ice.
At the front of the track, we observe a cross section of the
unscattered \chv wavefront, followed by a diffuse light cloud as the
photons are scattered away from the geometrical \chv cone. At depths
with higher dust concentrations, photons are obstructed by scattering
and absorbed before they can travel very far. This deforms the conical
light front, which appears to be bent backwards.
In the dusty region near $z=-350$~m, the photon flux is depleted and the
surviving photons delayed by increased scattering. Muon track
reconstruction is often strongly dependent on the earliest recorded
photons, corresponding to the \chv wavefront. Distortions in the
wavefront, like in Fig. \ref{infmuonpic}(b), can degrade the
reconstruction accuracy unless they are taken into account by the track
fitting algorithms.

\begin{figure*}[htb]
\center
\includegraphics[width=\textwidth]{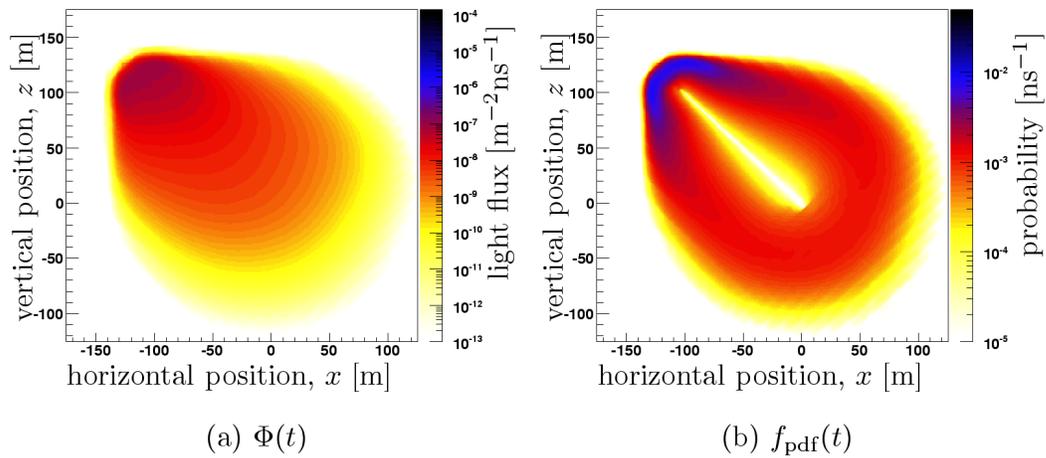}
\caption{
\label{muonpic}
Light distribution from a 142~m long muon track without secondary interactions.
This muon was created at the origin and propagated upward at
$\Theta_{\src}=135$\degr~until it decayed at $(x,z)=(-100,100)$.
The figure shows a snapshot 147~ns after the muon disappeared.
Both the photon flux (a) and the probability density function (b)
for such a comparably short track are similar to those
produced by a point-like cascade.
}
\end{figure*}

Fig.~\ref{muonpic} shows a snapshot of the light field generated by a finite muon
track without secondary interactions. The muon was created at the origin and propagated upward at
$\Theta_{\src}=135$\degr~until it decayed after 142~m. The probability density function for such a relatively short track
approaches a shape similar to that of point-like cascades, making it hard to distinguish the two
cases in an experimental situation with a limited number of points sampled by light sensors.
\px{}-based simulations with realistic medium and light source descriptions allow experimentalists
to isolate the differences
in light profiles for these (and other) distinct cases and to develop appropriate reconstruction algorithms.

\clearpage

\section{Conclusion}

In this paper we have presented the concepts and methods
which combine into the \px{} software package. We have explained how the
program can be used for calculating and tabulating light distributions
around a stationary or moving source, as a function of time and space in
scattering and absorbing heterogeneous media. The light distributions
obtained from our Monte Carlo simulation agree well with observations of
calibration light sources in deep sea water and glacial ice surveys. In
the last section it is demonstrated how \px{} can be used to model how
optical inhomogeneities of the Antarctic ice at the location of the
IceCube neutrino telescope distort the footprints of elementary particle
interactions.

\section{Acknowledgments}

We are grateful to Dr.\ Nathalie Palanque-Delabrouille for supplying
data from the {\sc{Antares}} water surveys and her helpful
advice on the implementation of water specific parameters.
We also thank the members
of the \amanda and IceCube collaborations for fruitful discussions and
useful feedback.

\appendix

\end{document}